\documentclass[apj]{emulateapj}
\usepackage{amsmath}

\begin{document}
\title{Rapidly decaying supernova 2010X: a candidate ``.Ia'' explosion}

\author{Mansi M. Kasliwal\altaffilmark{1}, S. R. Kulkarni\altaffilmark{1}, Avishay Gal-Yam\altaffilmark{2}, Ofer Yaron\altaffilmark{2}, Robert M. Quimby\altaffilmark{1}, Eran O. Ofek\altaffilmark{1}, Peter Nugent\altaffilmark{3}, Dovi Poznanski\altaffilmark{3,12}, Janet Jacobsen\altaffilmark{3}, Assaf Sternberg\altaffilmark{2}, Iair Arcavi\altaffilmark{2}, D. Andrew Howell\altaffilmark{4,16}, Mark Sullivan\altaffilmark{5}, Douglas J Rich\altaffilmark{6}, Paul F Burke\altaffilmark{7}, Joseph Brimacombe MB ChB FRCA MD\altaffilmark{8,9}, Dan  Milisavljevic\altaffilmark{10}, Robert Fesen\altaffilmark{10}, Lars Bildsten\altaffilmark{11,16}, Ken Shen\altaffilmark{12}, S. Bradley Cenko\altaffilmark{12}, Joshua S. Bloom\altaffilmark{12}, Eric Hsiao\altaffilmark{3}, Nicholas M. Law\altaffilmark{13}, Neil Gehrels\altaffilmark{14}, Stefan Immler\altaffilmark{14}, Richard Dekany\altaffilmark{15}, Gustavo Rahmer\altaffilmark{15}, David Hale\altaffilmark{15}, Roger Smith\altaffilmark{15}, Jeff Zolkower\altaffilmark{15}, Viswa Velur\altaffilmark{15}, Richard Walters\altaffilmark{15}, John Henning\altaffilmark{15}, Kahnh Bui\altaffilmark{15} \& Dan McKenna\altaffilmark{15}}

\altaffiltext{1}{Cahill Center for Astrophysics, California Institute of Technology, Pasadena, CA, 91125, USA}
\altaffiltext{2}{Benoziyo Center for Astrophysics, Faculty of Physics, The Weizmann Institute of Science, Rehovot 76100, Israel}
\altaffiltext{3}{Computational Cosmology Center, Lawrence Berkeley National Laboratory, 1 Cyclotron Road, Berkeley, CA 94720, USA}
\altaffiltext{4}{Las Cumbres Observatory Global Telescope Network, Inc, Santa Barbara, CA, 93117, USA}
\altaffiltext{5}{Department of Physics, Oxford University, Oxford, OX1 3RH, UK}
\altaffiltext{6}{Rich Observatory, 62 Wessnette Dr., Hampden, Maine USA}
\altaffiltext{7}{Burke Observatory, 19 Berry Rd., Pittsfield, Maine USA}
\altaffiltext{8}{New Mexico Skies Observatory, Mayhill New Mexico, USA}
\altaffiltext{9}{James Cook University, Cairns, Australia}
\altaffiltext{10}{6127 Wilder Lab, Department of Physics and Astronomy, Dartmouth College, Hanover, NH, 03755, USA}
\altaffiltext{11}{Kavli Institute of Theoretical Physics, University of California Santa Barbara, Santa Barbara, CA 93106, USA}
\altaffiltext{12}{Department of Astronomy, 601 Campbell Hall, University of California, Berkeley, CA 94720-3411, USA}
\altaffiltext{13}{Dunlap Institute for Astronomy and Astrophysics, University of Toronto, 50 St. George Street, Toronto M5S 3H4, Ontario, Canada}
\altaffiltext{14}{NASA-Goddard Space Flight Center, Greenbelt, MD 20771, USA}
\altaffiltext{15}{Caltech Optical Observatories, California Institute of Technology, Pasadena, CA 91125, USA}
\altaffiltext{16}{Department of Physics, University of California Santa Barbara, Santa Barbara, CA 93106, USA}

\begin{abstract}
We present the discovery, photometric and spectroscopic follow-up observations
of SN\,2010X (PTF\,10bhp). This supernova decays exponentially with $\tau_d=$\,5\,days, 
and rivals the current recordholder in speed, SN\,2002bj. SN\,2010X peaks at M$_r=-$17\,mag 
and has mean velocities of 10,000 km s$^{-1}$. Our light curve modeling suggests
a radioactivity powered event and an ejecta mass of 0.16 M$_{\odot}$. If powered by Nickel, 
we show that the Nickel mass must be very small 
($\approx$0.02 M$_{\odot}$) and that the supernova quickly becomes optically thin 
to $\gamma$-rays. Our spectral modeling suggests that SN\,2010X and SN\,2002bj have
similar chemical compositions and that one of Aluminum or Helium is present.  If 
Aluminum is present, we speculate that this may be an accretion induced collapse of 
an O-Ne-Mg white dwarf. If Helium is present, all observables of SN\,2010X 
are consistent with being a thermonuclear Helium shell detonation on a white dwarf, 
a ``.Ia'' explosion. With the 1-day dynamic-cadence experiment on the Palomar Transient
Factory, we expect to annually discover a few such events.  
\end{abstract}
\keywords{supernovae: individual (SN\,2010X, SN\,2002bj), white dwarfs, surveys}

\section{Introduction}
\label{sec:Introduction}

Our present knowledge of cosmic explosions is arguably biased by
the searches themselves. In particular, the cadence and depth of many supernovae
searches are designed to efficiently discover supernovae of type
Ia (SNe\,Ia). A repeat visit to the sky on timescales of
five days maximizes sky coverage and is still sufficient to catch SNe\,Ia
on the rise.  The brilliance of these events, peak absolute visual magnitude of $-19$,
sets the sensitivity of the searches. Conversely, fainter events and
those with a shorter characteristic lifetime are likely to be missed in 
such searches.

\begin{figure*}[htbp]
    \centering
    \includegraphics[width=3.5in,height=2.9in]{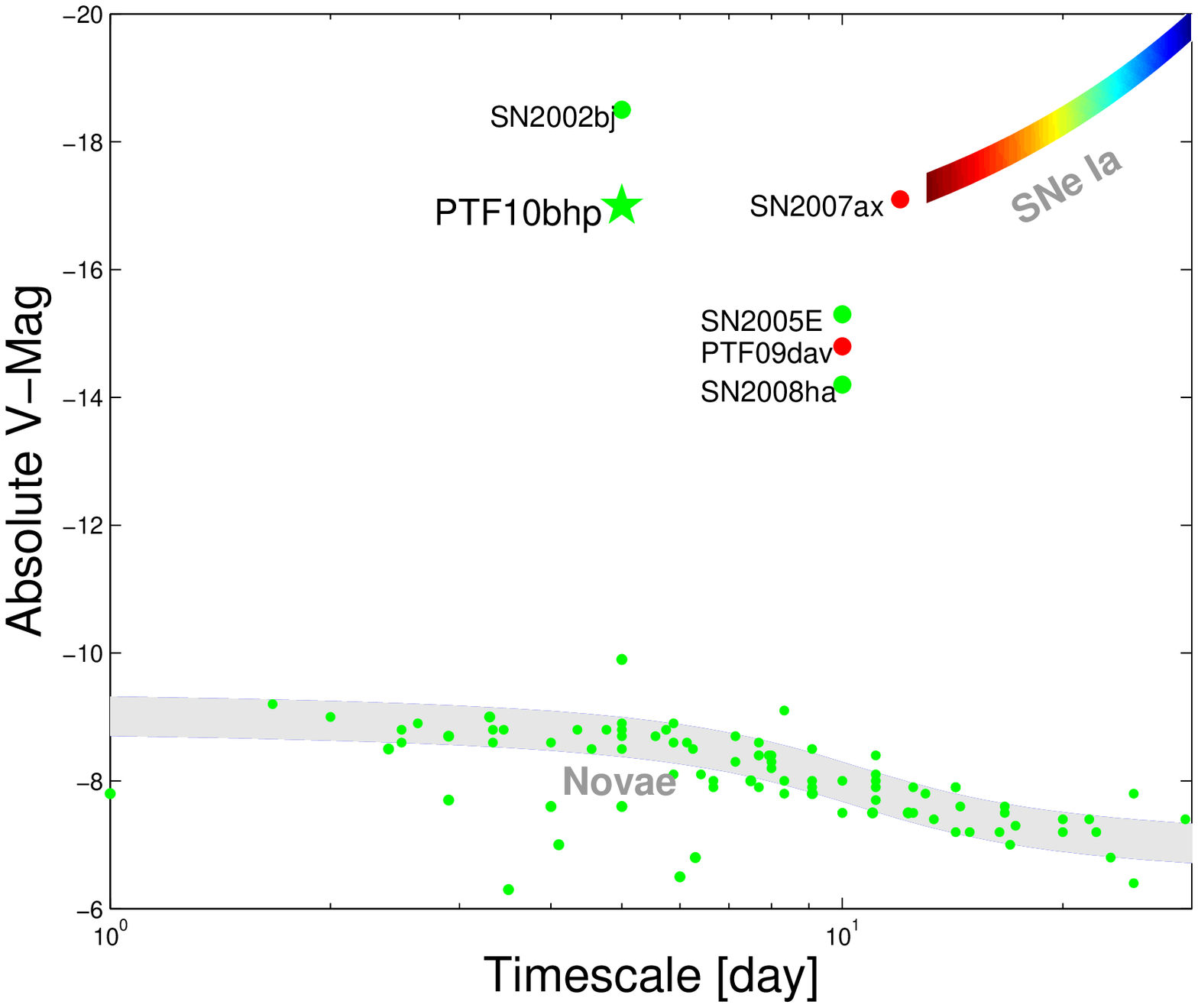}\includegraphics[width=3.5in,height=2.9in]{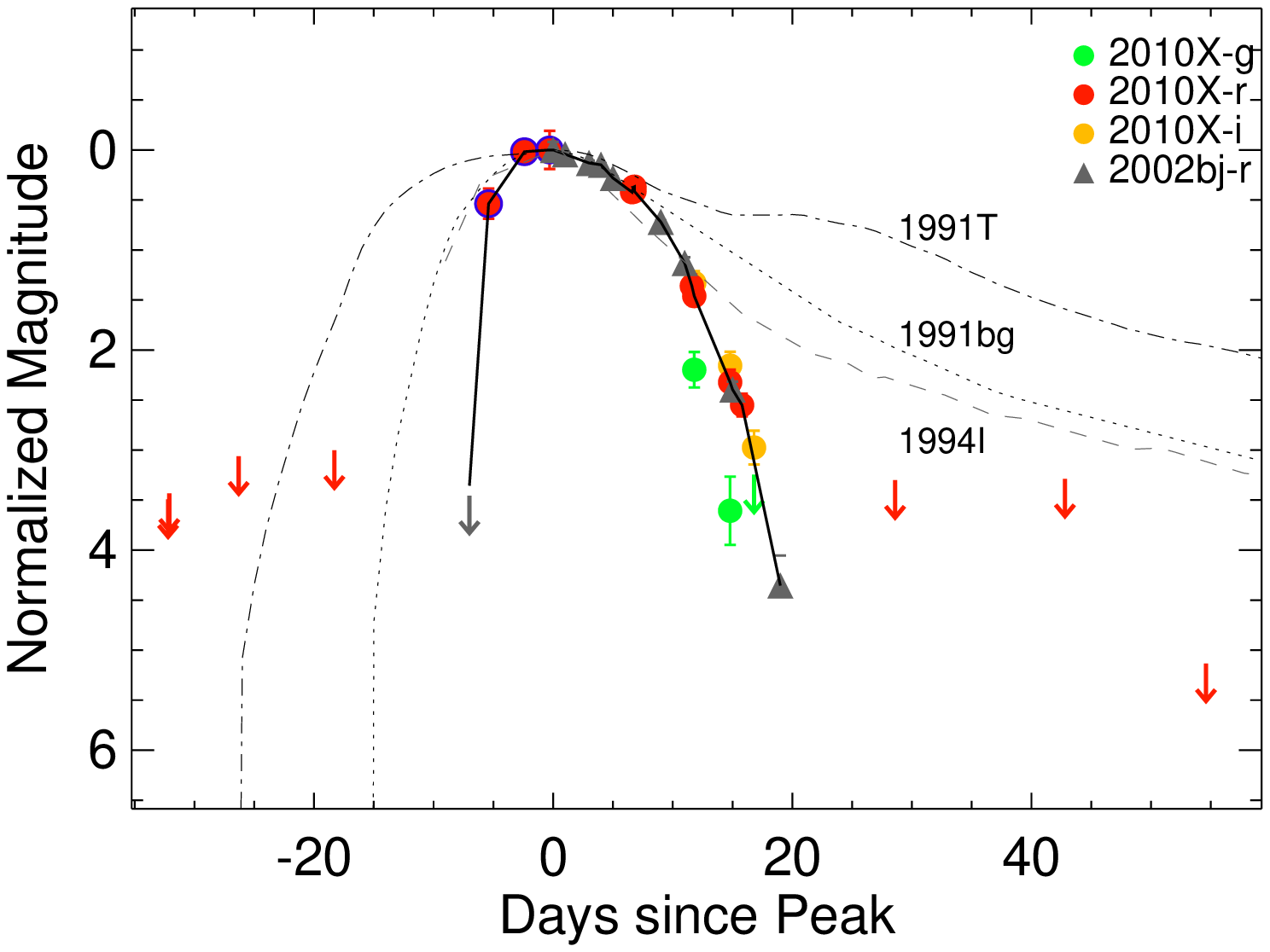}
    \caption[]{\small
    {\it \bf Left:} The phase space of cosmic explosive transients. The color
    for each event represents the color at peak brightness.
    The band to the top right denotes supernovae of type Ia. The fastest such
    event is SN\,2007ax \citep{kog+08}. Classical novae occupy a
    band between $-6$ and $-10$ magnitude. Note that the only two
    transients with a timescale shorter than ten days are SN\,2010X
    (PTF\,10bhp) and SN\,2002bj. 
    {\it \bf Right:} The multi-band optical light curve of SN\,2010X (colored circles;
    green is g-band, red is r-band, orange is i-band). Three white
    light measurements have been calibrated to r-band and denoted
    by red circles with blue outline. Downward arrows represent upper limits.
    All light curves are normalized and shifted so that peak magnitude 
    is zero and the time at peak is set to zero. For SN\,2010X the 
    epoch of maximum light is at MJD of 55239.5.
    The fast evolution of SN\,2010X is compared to the current recordholder
    for fast supernovae, SN2002bj (gray triangles; r-band; \citealt{pcn+10}). 
    Also shown is a prototypical ``fast'' Type Ic supernova, SN1994I 
    (dashed line; \citealt{rvh+96}) and templates 
    {\footnote{http://supernova.lbl.gov/\textasciitilde nugent/nugent\_templates.html}}
    of the fast Type Ia SN1991bg and slow Type Ia SN1991T \citep{nkp02}. Note the rapid rise 
    and the spectacular decay of SN\,2010X and SN\,2002bj relative to the
    other Type I exemplars.
}
    \label{fig:taumv}
\end{figure*}

To illustrate the unexplored nature of this phase space, we plot
the luminosity of 
optical transients versus their characteristic timescale
(Figure~\ref{fig:taumv}). SNe\,Ia are confined to a narrow
band (\citealt{p93}) with decay timescales ranging
from twelve days to three weeks.  Classical novae span a large range
of timescales albeit at considerably lower luminosities.

Figure~\ref{fig:taumv} brings two white-spaces to attention: the wide
``gap'' in luminosity between novae and supernovae, and 
the apparent paucity of luminous events on short timescales. 

Next, we discuss currently known exemplars of ``faint'' 
(i.e. lower luminosity than SNe\,Ia)  and ``fast'' (i.e. faster than
SN\,2007ax) transients. SN\,2005E occurred in the halo of its host galaxy and 
has been proposed as a Helium detonation on a binary white dwarf \citep{pgm+10}.
SN\,2005cz has been proposed to have a massive star origin \citep{kmn+10}. 
SN\,2008ha is also being widely debated both 
as a deflagration of a white dwarf \citep{fcf+09,fbr+09} and  
core-collapse of a massive star \citep{vpc+09}. 

Until recently, the fastest event known was SN\,2002bj \citep{pcn+10}. 
It decayed by one magnitude in five days and was quite spectroscopically 
peculiar. The origin of this event is not yet clear.  

The Palomar Transient Factory{\footnote{http://www.astro.caltech.edu/ptf}} 
(PTF) was motivated
in great measure to systematically explore the phase space for 
fast and faint explosive transients \citep{lkd+09,rkl+09}.  Here, 
we present the discovery of a fast event, SN\,2010X (PTF\,10bhp).

\section{Discovery}
\label{sec:discovery}

On UT 2010 February 7.07, D. Rich of Hampden, Maine discovered a
transient in the galaxy NGC\,1573A at RA(J2000)=04$^{\rm h}$48$^{\rm m}$27.7$^{\rm s}$
and Dec(J2000)=+73$^{\circ}$28$\arcmin$13$\arcsec$. The discovery was confirmed by
P. Burke of Pittsfield, Maine, upon which a notification was issued
(CBET 2166; \citealt{rb10}) and the transient dubbed SN2010X.
On UT 2010 February 19.13, the Palomar Transient Factory independently
detected this same transient and the pipeline assigned the name, PTF\,10bhp. 
PTF had previously undertaken observations of this field (as a part of the
dynamic cadence experiment) on January 11, 17 and 25 but with no
detection.

\section{Optical Light Curve}
\label{sec:lc}

Energized by the apparent rapid fading, we initiated follow-up
observations.  The photometric observations from the 2-m Faulkes
North Telescope (FTN) of the Las Cumbres Observatory Global Telescope
(LCOGT), PTF, the Palomar Hale 200-inch telescope (P200) as well
as white light observations provided by our amateur astronomer
colleagues are summarized in Figure~\ref{fig:taumv}. 

SN\,2010X is located close to the nucleus of its host galaxy 
(4.4\arcsec\,E, 6.0\arcsec\,N) and as such
galaxy light subtraction is critical to produce reliable
photometry.  The images were subtracted from a template image using
the software \texttt{hotpants} and \texttt{wcsremap} to measure a
convolution kernel and align the images respectively (both codes
supplied by A. Becker{\footnote
{http://www.astro.washington.edu/users/becker/c\_software.html}}). Aperture
photometry was performed on each of these in a self-consistent
manner using the same set of 22 calibration stars.  Conversions
from USNO-B1 magnitudes to SDSS gri magnitudes were done adopting
\citet{jga06}. The resulting light curve is plotted in Figure~\ref{fig:taumv}.

Overplotting SN\,2002bj, we find that light curves of the two supernovae are
remarkably similar. Linearly fitting all the r-band detections post
maximum light, we measure that SN\,2010X decayed by $0.23\pm0.01$\,
mag day$^{-1}$.  The corresponding exponential timescale (in the
$r$-band) is $\tau_d= 4.7\pm 0.2$\,days.

The foreground Galactic extinction along the line of sight is E($B-V$)=0.146 
or A$_r$=0.4 \citep{sfd98}. The redshift of NGC~1573A is 0.015. Assuming standard 
cosmology (and $h_{0}$=0.72), we adopt a distance of 62.5\,Mpc and a distance modulus 
of 34.0. Thus, the peak absolute magnitude of SN\,2010X is $M_{r}\approx\,-$17.0\,mag, 1.5\,mag less 
luminous than SN2002bj.

\section{Spectroscopic Follow-Up}
\label{sec:spec}

\begin{figure}[htbp]
   \centering
   \includegraphics[height=3.5in,width=3.5in]{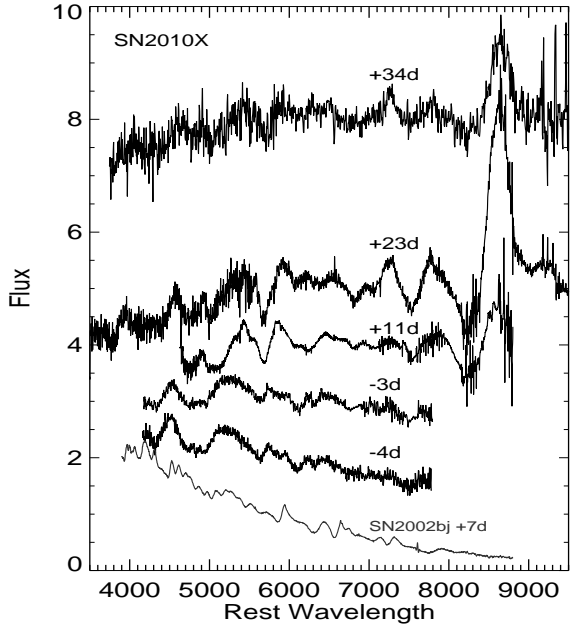} 
   \caption{Spectroscopic follow-up of SN\,2010X by MDM, Gemini,
   Keck and Palomar Observatories. The phase of the spectra relative
   to maximum light is labeled. Note the velocity evolution. Also shown
   is a spectrum of SN\,2002bj \citep{pcn+10}. 
}
   \label{fig:series}
\end{figure}

On February 8 and 9, the first spectra (Figure~\ref{fig:series}) to classify the
nature of this transient were taken with CCDS on the 2.4\,m Hiltner
telescope of the MDM observatory (CBET 2167, \citealt{mf10}).
Comparison with a library of supernova spectra using SNID \citep{bt07}
showed resemblance to the Type Ic supernovae SN\,1994I and SN\,2004aw
a few days before maximum light. Further observations (Figure~\ref{fig:series}) were undertaken 
on Gemini-North/GMOS (Feb 23), Keck~I/LRIS (Mar 7) and the Hale 200-inch/DBSP (Mar 18) 
telescopes. 
No perfect matches to Ic (or Ia, Ib) templates were found for
these spectra. The velocity evolved from 12000\,km s$^{-1}$ before 
maximum to 9000\,km s$^{-1}$ at late-time. 

\begin{figure}[htbp]
   \centering
   \includegraphics[height=3.5in,width=3.0in]{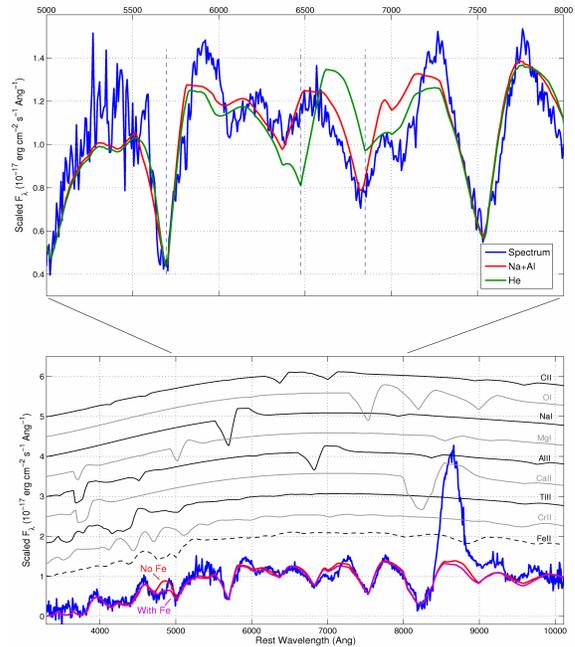} 
   \caption{SYNOW fit to the Keck spectrum (+23\,days) of SN\,2010X. Lines contributed by each ion 
    are shown. Fits with (purple) and without (red) Iron are overplotted on the data (blue).
    {\it \bf Top Panel: }The dilemma of whether SN\,2010X has Helium or a combination of Sodium and Aluminum. 
    The vertical dashed lines show Helium at 9500 km s$^{-1}$. In non-LTE, the singlet transition of 
    $\lambda$6678 may be suppressed relative to the $\lambda$7065 and $\lambda$5876 triplet transition 
    Helium lines.  
   } \label{fig:synow}
\end{figure}

We used SYNOW \citep{jb90} to infer elements in the spectra of SN\,2010X 
(Figure~\ref{fig:synow}). The most prominent identifications are oxygen 
(O I lines), Calcium (both Ca II IR triplet and Ca II H+K on the blue side), 
Carbon (C II lines),  Titanium (Ti II) and Chromium (Cr II). 
Ti II and Cr II explain the broad blue features and adding Fe II improves the
fit slightly. There is also some evidence for Mg I albeit based on single line.

The presence of Helium (He I), Sodium (Na D) and Aluminum (Al II) is less
clear and we illustrate this dilemma in the inset of Figure~\ref{fig:synow}. 
He I has three relevant lines: 5876\,\AA\ , 6678\,\AA\  and 7065\,\AA\,. The 
absorption feature around 5700\,\AA\ can be explained by both He I as well 
as Na D. The absorption feature around 6850\,\AA\ can be explained by Al II or
He I. Since the central He I line is not prominent, SYNOW suggests
that the combination of Na D and Al II is a better fit. However, \citet{b03}
discuss that this central He I line is a singlet transition
and this may both be suppressed and blueshifted in non-LTE relative to
the other two He I triplet transitions. Therefore, we cannot conclusively 
say whether or not Helium is present in SN\,2010X.

Comparing the spectra, SN\,2002bj has substantially lower velocities 
(4000 km s$^{-1}$ at +7d vs. 10,000 km s$^{-1}$ at +10d) and 
a bluer continuum ($g-r$=0.2 at +12d vs. $g-r$=1.2 at $+$23d) than SN\,2010X. 
Consistent with the SYNOW fit shown in \citealt{pcn+10}, 
the elements in common between the two supernovae are O I, C II and Mg II. 
The primary difference is the presence of Ca II in SN\,2010X and 
presence of S II in SN\,2002bj. We re-fit the spectrum of SN\,2002bj with 
the same elements as in SN\,2010X. We find that the presence of Al II vs. 
He I is just as ambiguous for SN\,2002bj as SN\,2010X. Similar to SN\,2010X,
including Fe II improves the fit but the presence of Fe-group elements 
in  SN\,2002bj is not conclusive.

\section{Modeling the light curve}
\label{sec:physics}

The excellent match between the normalized light curves of 
SN\,2010X and SN\,2002bj (see Figure~\ref{fig:taumv}) suggests 
that these two SNe belong to the same class of explosions. 
Combining the two data sets allows a
robust determination of the rise time{\footnote{time from explosion to peak brightness}} 
($\tau_r\approx 6\,$d) and the subsequent exponential decay ($\tau_d\approx 5\,$d)

The peak bolometric luminosity of SN\,2010X is $L_{\rm peak}=10^{42}\,{\rm
erg\,s}^{-1}$. While the expansion speed varies from 12,000\,km\,s$^{-1}$
at early times to 9,000\,km\,s$^{-1}$ at late times, we
accept $v_s\approx 10,000\,{\rm km\,s}^{-1}$ as a representative
value. 

The rise time in an explosion is the geometric mean of the initial
photon diffusion timescale and the initial hydrodynamic time scale
{\footnote{The derivation can be found in the textbooks, \citealt{a96}
and \citealt{p01}.}}. Thus,
${\tau_r}^{2} \propto \kappa M_{\rm ej}/v_s$ where $\kappa$ is the
opacity.  Assuming that the mean opacity of SN\,2010X is the
same as that for SNe\,Ia events, (for which, following
\citealt{hgk+10},
we adopt the following:
$M_{\rm ej}\approx 1.4\,M_\odot$, $v_s=10^9\,{\rm
cm\,s}^{-1}$ and $\tau_r\approx 17.5\,$d), we obtain
$M_{\rm ej}\approx 0.16\,M_\odot$.  This gives an explosion energy,
$E_0 = 1/2M_{\rm ej}v_s^2\approx 1.7\times 10^{50}$\,erg.

Next, we investigate a physical model that satisfactorily accounts
for the rise time, the decay time, the peak luminosity and the
expansion velocity.

\subsection{Pure explosion}

The simplest model is an explosion in
which all the explosive energy ($E_0$) is deposited instantaneously
into the ejecta.  The peak luminosity is then $E_0/t_d(0)$ where
$t_d(0) \propto \kappa M_{\rm ej}/R_0$ is the initial photon diffusion
time; here, $R_0$ is the radius of the progenitor.  Following
peak luminosity, the decay is rapid: $\log(L)\propto
-(t/\tau_r)^2$. The virtue of this model is that one can obtain an
arbitrarily rapid rate of decay since, over any limited stretch of
time, the light curve can be approximated by a linear decay with
the desired value for the slope.

For SN\,2010X, we find $R_0\sim 4\times 10^{12}\,$cm.  The
large inferred radius would make sense if the progenitor had an
envelope (as in type II supernovae).  The absence of hydrogen at
any phase of the supernova (see \S\ref{sec:spec}) argues strongly
against this model.  Hence, we reject this hypothesis.

\subsection{Radioactivity powered explosion}

The next level of models is that developed for SNe\,Ia explosions, where
the peak luminosity and subsequent decay is governed by radioactive
material present in the ejecta. In this model, expansion
decreases the store of internal energy  whereas radioactivity increases it.
If the photon diffusion time-scale is long, most of the radioactive
energy goes into expansion. Once the diffusion time-scale becomes
smaller than the expansion time-scale, the light curve tracks the
radioactive luminosity \citep{a82}, provided that there is
sufficient optical depth for the $\gamma$-rays emitted during
radioactive decay to undergo multiple scatterings and lose their
energy to electrons.

The primary source of luminosity in a SN\,Ia model is the heat
provided by $\gamma$-rays emitted as $^{56}$Ni decays to
$^{56}$Co and then to $^{56}$Fe. In SNe\,Ia, the column
density of the ejecta is thick enough to trap the $\gamma$-rays and
successive Compton scatterings extract energy from the $\gamma$-rays
(at least for the first month).  However, given the small
ejecta mass for SN\,2010X, attention has to be paid to the possibility
that $\gamma$-rays from decaying nuclei may escape without depositing
their energy into the ejecta.

The electron (Thompson) optical depth is: 
\begin{eqnarray}
\tau_{e} & =& n_e R \sigma_{T}  = \frac{3}{4\pi} \frac{M_{\rm ej}}{m_p} \frac{Z}{A} \frac{\sigma_{T}}{R^2} \cr
             & \sim& 9\left(\frac{M_{\rm ej}}{0.16 M_{\odot}}\right) \left(\frac{t}{15\,{\rm day}}\right)^{-2} 
\end{eqnarray}
where $Z$ is the atomic number, $A$ is the mass number, m$_p$ is
mass of proton, $\sigma_T$ is the Thompson cross-section and $R\sim
{\rm 6(t/day)\,AU}$ is the radius at time $t$.

Thus, there appears to be sufficient optical depth at the epoch
of peak luminosity to trap most of the $\gamma$-rays.  Thus, for
SN\,2010X, the peak luminosity of $10^{42}\,{\rm erg\,s}^{-1}$
corresponds to $^{56}$Ni mass of about $0.02\,M_\odot$ -- a very
small amount by the standards of most supernovae. For SN\,2002bj,
the peak luminosity was $10^{43}\,{\rm erg\,s}^{-1}$ \citep{pcn+10}
and the inferred $^{56}$Ni mass was correspondingly larger, $0.2\,M_\odot$.

\begin{figure}[htbp]
    \centering
    \includegraphics[width=3.5in]{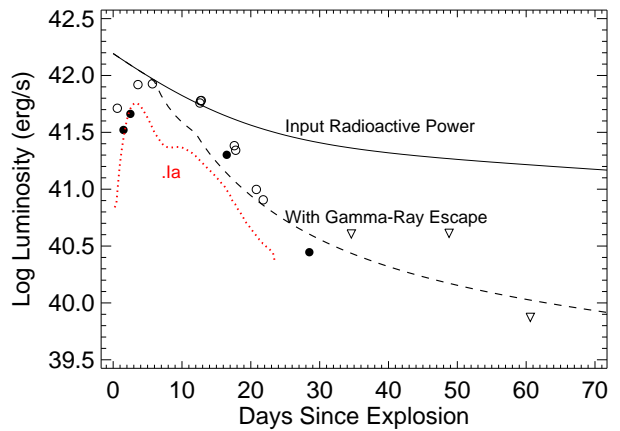} 
\caption{\small
Shown above is the radioactive luminosity (solid line)  and
absorbed luminosity (dashed line) for the following model
parameters: $M_{\rm ej}=0.16\,M_\odot$, $M_{\rm Ni}=0.02\,M_\odot$
and $v=10^9\,{\rm cm\,s}^{-1}$. 
Also shown is a quasi-bolometric light curve of SN\,2010X estimated 
by (a) computing $\nu$F$_{\nu}$ in $r$-band 
(empty circles are detections, inverted triangles are upper limits), 
and (b) integrating the optical spectrum (filled circles). Also shown
is a comparison to a ``.Ia'' light curve (red dotted line; \citealt{skw+10})
assuming: M$_{\rm wd}$=1.2 M$_{\odot}$, M$_{\rm env}$=0.05 M$_{\odot}$,
M$_{\rm ej}$=0.036 M$_{\odot}$, M$_{\rm Fe}$= 0.005 M$_{\odot}$, M$_{\rm Ni}$=0.02 M$_{\odot}$, M$_{\rm Cr}$= 0.0002 M$_{\odot}$.
} 
\label{fig:bollc} 
\end{figure}

Next, we use a fitting formula (as given in \citealt{k05}; Equation
47) to estimate the fraction of $\gamma$-rays which are effectively
absorbed inside the ejecta, $\eta(\tau_e)$. The kinetic energy of 
positrons (3.5\% of $L_{\rm Co}$; \citealt{shc+02}) dominates by day 51. 
Hence, the radiated luminosity, $L_{\rm rad}=(0.965\eta + 0.035) L_{\rm Co} + \eta L_{\rm Ni}$ 
where $L_{\rm Ni}$ is the radioactive power released by the decay of $^{56}$Ni, 
and  $L_{\rm Co}$ by the daughter $^{56}$Co. In Figure~\ref{fig:bollc}, we display 
the luminosity due to radioactivity and that actually trapped in the ejecta   
--- the latter shows a satisfactory agreement with the observations.

\subsubsection{Possible X-ray signature}
An optically thin ejecta opens up the possibility of detecting 
the $\gamma$-rays (or degraded hard X-rays) emitted during $\beta$-decay.
The {\it Swift} Observatory observed SN\,2010X for 9758.7\,s on  MJD  55248.775
(9 days past peak). We constrain the X-ray flux
{\footnote{we note that the six photons in the XRT HPD PSF of 0.3\arcmin 
are likely from the galaxy nucleus}} to be less than 0.00050 counts s$^{-1}$ or 
7.7$\times$10$^{39}$ erg s$^{-1}$. By this epoch, our model shows that 
$L_\gamma\sim 10^{41}\,{\rm erg\,s}^{-1}$. Since photon number
is conserved in scattering, the luminosity in the {\it Swift} band 
is expected to be a factor of $200$ smaller and hence, the
upper limit is not constraining. 

\section{Environment}
\label{sec:envrnmt}
\begin{figure}[htbp]
   \centering
   \includegraphics[width=3.5in]{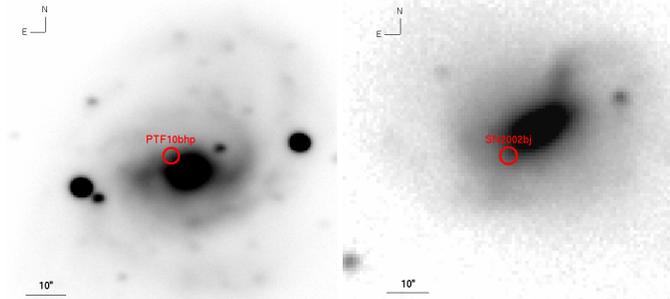} 
   \caption[]{{\it Left:} R-band image of NGC~1573A, the host of SN\,2010X,
 taken with the Large Format Camera on the Palomar 200-inch telescope. 
{\it Right:} Sum of all available Deepsky
{\footnote{http://supernova.lbl.gov/\textasciitilde nugent/deepsky.html.}}
\citep{n09} images of NGC1821, the host of SN2002bj.
}
   \label{fig:hosts}
\end{figure}

The host of SN\,2010X, NGC~1573A, is a small (1.6\arcmin~diameter),
spiral galaxy variously classified as Sb (UGC) and SABbc (RC3).
The host of SN\,2002bj, NGC~1821, is a small (1.1\arcmin~diameter), 
barred irregular galaxy classified as IB(s)m. Both transients occurred 
close to the galaxy nucleus --- 2.3\,kpc for SN\,2010X and 1.8\,kpc for SN\,2002bj. 
In Figure~\ref{fig:hosts}, we show the location of the supernovae in deep images of the galaxy.

\section{Conclusion}
To summarize, SN\,2010X is the second member of a class of
supernovae that declines exponentially on timescales shorter than 5\,days.
Relative to SN\,2002bj, SN\,2010X is less luminous by 1.5\,mag (M$_R \approx\,-$17) 
and has higher velocities (10,000 km s$^{-1}$) by more than a factor of two.
Both events have a small inferred ejecta mass. 
Both events are spectroscopically different from any other type I supernovae.
The spectra for both supernovae can be modeled with mostly similar elements  
(C, O, Mg, Si, Ti and Fe). The evidence (or lack thereof) for Helium
is not conclusive in both cases.

If SN\,2010X is powered by radioactive $^{56}$Ni, the combination
of a rapid rise time and low peak luminosity constrains the Nickel
mass to be small, $0.02\, M_{\odot}$.  $^{56}$Ni constitutes $\approx$13\%
of ejecta. However, under the same assumptions, $^{56}$Ni would
constitute bulk of the ejecta mass for SN\,2002bj.  Thus, while in
both cases the ejecta mass remains the same, the nucleo-synthesis
may be strongly variable. We also show that given the small ejecta mass,
$\gamma$-rays from decaying $^{56}$Ni can start escaping from the
ejecta shortly after peak brightness.  This early escape reasonably
accounts for the rapid decay of the light curve of SN\,2010X.

\citet{pba+10} have argued that S\,Andromeda (the first recorded
SN in Andromeda) and SN\,1939B (the first recorded SN in Virgo) 
are also like SN\,2002bj. The claim primarily rests on rapid rise and 
rapid decay at early-time. It is of some interest to note that the late-time
(2 months to nearly 1 year) decay rates, 0.03 mag/day for S\,Andromeda 
and 0.02 mag/day for SN\,1939B \citep{pba+10}, are consistent with $^{56}$Co decay
(with some escape of $\gamma$-rays). A consistent explanation would require
a two-zone model: comparable amount of $^{56}$Ni in a slowly expanding
core (to account for the late time light curve) 
and a rapidly expanding shell (to account for the rapid decay
seen after peak brightness).

An alternative model is that the early-time emission is powered by
another suitably rapidly decaying radio-active element(s).
If powered solely by $^{48}$Cr, 0.02 M$_{\odot}$ is adequate. 
Recently, \citealt{skw+10} computed models
and observables for ``.Ia'' explosions \citep{bsw+07} powered by
$^{48}$Cr, $^{52}$Fe and $^{56}$Ni: rise time between 2--10\,days, 
ejecta velocity between 9000--13000 km s$^{-1}$, peak luminosity between 
0.5--5$\times$10$^{42}$ erg s$^{-1}$ and presence of Ca II and Ti II in the spectra. 
The properties of SN\,2010X are consistent with all these predictions. Specifically,
the light curve model presented for a core mass of 1.2 M$_{\odot}$ and envelope mass
of 0.05 M$_{\odot}$ is a reasonable match (Figure~\ref{fig:bollc}). 
Furthermore, \citealt{skw+10} also discuss that the presence of Helium in the 
spectra may be a non-LTE effect.   

If Aluminum is indeed present in the spectra, the avenue for
a speculative scenario opens up. Neither $^{26}$Al nor $^{27}$Al 
is a product of Helium burning. Aluminum can be made via explosive 
burning of Neon and/or Carbon \citep{ab97,ww80}. 
Perhaps, SN\,2010X is the outcome of accretion induced collapse 
of an O-Ne-Mg white dwarf \citep{mpq+09}.  

Finally, we note that the rich, star-forming environment of SN\,2010X 
and SN\,2002bj does not preclude a massive star origin. Fallback events, where
a massive star collapses into a black hole, are also expected
to be fast declining \citep{fbb+09,mtt+10}. However, the velocities expected 
from these models are significantly lower than observed and the spectra 
are more substantially dominated by intermediate mass elements. 

Regardless of all these rich possibilities, it is clear that further
progress in understanding the nature of these ephemeral transients would 
require a larger sample. Fortunately, PTF, especially as it moves to
``dynamic'' 1-day cadence \citep{lkd+09} targetting nearby galaxies and clusters, 
is well equipped to annually find a few such events.
Late-time photometry is important to look for tell-tale signatures of 
$^{56}$Co decay. Sensitive optical (or better still, ultra-violet) 
spectroscopy may directly reveal the radioactive element(s) powering 
these events. It is also not inconceivable (given the history of S\,Andromeda
and SN\,1939B) that we will be lucky enough to observe a local analog
of such an event with the hard X-ray mission, NuSTAR \citep{hbc+10}.


\smallskip
\noindent{\bf Acknowledgments.} 
MMK thanks the Gordon and Betty Moore Foundation for a Hale Fellowship
in support of graduate study. MMK thanks the Pumarth Headquarters
in Indore, India for their warm hospitality while writing this
manuscript.

We thank the referee for accepting this paper on a timescale
faster than that characterizing the rapid photometric evolution of this 
supernova!

We would like to acknowledge the following discussions: 
MMK \&\ Paolo Mazzali, AGY \&\ David Branch, SRK \&\ Xiaofeng Wang, 
DAH \&\ Ryan Foley, MMK \&\ Rollin Thomas. 
We are grateful to the staff of the Gemini Observatory  and Swift Observatory
for efficiently executing TOO triggers. We thank the librarians who maintain the 
ADS, the NED, and SIMBAD data systems. 

The Weizmann Institute PTF participation is supported by grants
to AGY from the Israel Science Foundation and the US-Israel Binational
Science Foundation. EOO and DP are supported by an Einstein Fellowship.
SBC is grateful for support from Gary and Cynthia Bengier and the Richard 
and Rhoda Goldman Fund. 
Computational resources and data storage were contributed by NERSC, supported by 
U.S. DoE contract DE-AC02-05CH11231. PEN acknowledges support from the US DoE 
contract DE-FG02-06ER06-04.

{\it Facilities:} \facility{PO:1.2m ()}, \facility{Hale ()}, 
\facility{Gemini:Gillett (),\facility{Hiltner ()},\facility{Swift ()}}


\end{document}